\documentstyle[aps,epsfig,amsmath,amssymb]{revtex}

\draft

\begin{document}

\title{Overcritical Rotation of a Trapped Bose-Einstein Condensate}

\author{A.~Recati$^a$, F.~Zambelli$^b$, and S.~Stringari$^b$}
\address{$^a$International School of Advanced Studies, Via Beirut 2/4,
I-34014 Trieste, Italy}
\address{$^b$Dipartimento  di Fisica, Universit\`a di Trento,}
\address{and Istituto Nazionale per la Fisica della Materia, 
I-38050 Povo, Italy}
\date{\today}

\maketitle

\begin{abstract}
The rotational motion of an interacting Bose-Einstein condensate
confined by a harmonic trap is investigated by solving the
hydrodynamic equations of superfluids, with the irrotationality constraint
for the velocity field. We point out the occurrence of an overcritical
branch where the system can rotate with angular velocity larger than
the oscillator frequencies. We show that in the case of isotropic trapping
the system exhibits a bifurcation from an axisymmetric to a triaxial
configuration, as a consequence of the interatomic forces.
The dynamical stability of the rotational motion with respect to
the dipole and quadrupole oscillations is explicitly discussed. 
\end{abstract}


An important peculiarity of harmonic trapping is the existence of a
critical angular velocity, fixed by the oscillator frequencies, above
which no system can rotate in conditions of thermal equilibrium.   
The main purpose of this work is to show that a trapped Bose-Einstein 
condensate at very low temperature can rotate at angular velocities
higher than the oscillator frequencies in a regime of dynamical equilibrium.  
The occurrence of overcritical rotations is a rather well established feature
in classical mechanics (see, for example, \cite{Lamb,Genta}) and is
the result of the crucial role played by the Coriolis force.
It is therefore interesting to understand the new features exhibited
by rotating superfluids and in particular the role played by
Bose-Einstein condensation.
The rotational behaviour of superfluids is, in fact, deeply influenced by 
the  constraint of irrotationality which makes it impossible, for such
systems, to rotate in a rigid way. 
Spectacular consequences of irrotationality are the quenching of the
moment of inertia with respect to the rigid value and the occurrence of 
quantized vortices \cite{Donnelly}. 
Both these effects have been recently observed in dilute gases
confined in harmonic traps \cite{Jila,Dalibard,Marago}.

We start our analysis by considering a dilute Bose gas interacting
with repulsive forces at zero temperature. 
For large systems, where the Thomas-Fermi approximation
applies, the equations of motion are well described by the
so called hydrodynamic theory of superfluids \cite{SS,RMP}.
If the confining potential rotates with angular velocity $\Omega$  
it is convenient to write these equations in the frame 
rotating with the angular velocity of the trap, where they take the form:
\begin{align}
&\frac{\partial \rho}{\partial t}+{\boldsymbol\nabla}
\big(\rho({\bf v}-{\boldsymbol\Omega}\times{\bf r})\big)=0
\label{Rcontinuity}\\
&\frac{\partial{\bf v}}{\partial
  t}+{\boldsymbol\nabla}\cdot\left(\frac{v^2}{2}+\frac{V_{\rm
  ext}({\bf r})}{M} +\frac{\mu_{\rm loc}(\rho)}{M}-{\bf
  v}\cdot({\boldsymbol\Omega}\times{\bf r})\right)=0\;. 
\label{Rveq}
\end{align}
In the above equations $V_{\rm ext}({\bf
r})=M(\omega_x^2x^2+\omega_y^2y^2+\omega_z^2z^2)/2$ 
is the oscillator potential providing the external confinement. Notice
that in the rotating frame $V_{\rm ext}({\bf r})$ does not depend on
time. Furthermore $\mu_{loc}(\rho)=g\rho$ is the chemical potential of
the uniform gas, where $g=4\pi\hslash^2a/M$ is the coupling constant
fixed by the $s$-wave scattering length $a$ and ${\bf v}$ is the
velocity field in the laboratory frame, expressed in terms of the coordinates 
in the rotating frame. It satisfies the irrotationality
constraint. Eqs. (\ref{Rcontinuity})-(\ref{Rveq}) can be applied also
to a trapped Fermi superfluid where the expression for
$\mu_{loc}(\rho)$ takes, of course, a different form.
The stationary solutions in the rotating frame are obtained by
imposing the conditions $\partial\rho / \partial t=0$ and 
$\partial{\bf v}/\partial t=0$. 
Let us look for solutions of the form \cite{nota}
\begin{equation}
{\bf v}=\alpha
{\boldsymbol\nabla}(xy)
\label{virr}
\end{equation}
for the velocity field, where $\alpha$ is a parameter that will be
determined later. 
Choice (\ref{virr}) for the velocity field rules out
the description of vortical configurations. Vortices cannot
be in any case described by the hydrodynamic equations
(\ref{Rcontinuity})-(\ref{Rveq}) since they require the use of
more microscopic approaches, like Gross-Pitaevskii theory for the order
parameter,
accounting for the  behaviour of the system at distances of
the order of the healing length \cite{Lev,RMP}. Since the critical frequency needed to
generate a stable vortex becomes smaller and smaller as the number of atoms
increases \cite{Franco}, the solutions discussed in the present work
correspond, in general, to metastable configurations.

By substituting expression (\ref{virr}) into Eq. (\ref{Rveq}) one immediately 
finds that the resulting equilibrium density is given by the parabolic shape
\begin{equation}
\rho({\bf r})=\frac{1}{g}\left[\tilde{\mu} -
\frac{M}{2}(\tilde{\omega}_x^2x^2+\tilde{\omega}_y^2y^2+\omega_z^2z^2)\right] 
\label{Rdensity}
\end{equation}
also in the presence of the rotation. Of course Eq. (\ref{Rdensity})
defines the density only in the region where $\rho > 0$. Elsewhere one
should put $\rho =0$. The new distribution is characterized by the effective 
oscillator frequencies 
\begin{align} 
\tilde{\omega}^2_x 
&=\omega_x^2+\alpha^2-2\alpha\Omega
\label{tildex}\\
\tilde{\omega}^2_y 
&=\omega_y^2+\alpha^2+2\alpha\Omega\;,
\label{tildey}
\end{align} 
which fix the average square
radii of the atomic cloud through the relationships
\begin{equation}
\tilde{\omega}^2_x\langle x^2\rangle = \tilde{\omega}^2_y\langle
y^2\rangle = {\omega}^2_z\langle
z^2\rangle = {2\tilde{\mu} \over 7M}\;,
\label{radiimu}
\end{equation}
where the quantity
\begin{equation}
\tilde{\mu} =\frac{\hslash\tilde{\omega}_{ho}}{2}
\left (\frac{15Na}{\tilde{a}_{ho}}\right)^{2/5} 
\label{mu}
\end{equation}
is the chemical potential in the rotating frame and ensures  the
proper normalization of the density (\ref{Rdensity}). 
In Eq. (\ref{mu}) we have defined $\tilde{\omega}_{\rm ho}=
(\tilde{\omega}_x\tilde{\omega}_y\omega_z)^{1/3}$
and $\tilde{a}_{\rm ho}=\sqrt{\hslash/M\tilde{\omega}_{ho}}$. 
The applicability of the Thomas-Fermi approximation, and hence of 
the hydrodynamic equations (\ref{Rcontinuity})-(\ref{Rveq}), 
requires that the parameter $15Na/\tilde{a}_{\rm ho}$ be much larger
than unity. 
The rotation of the trap, providing a value of $\alpha$ different from
zero, has the consequence of modifying the shape of the 
density profile, through the change of the effective frequencies 
$\tilde{\omega}_x$ and $\tilde{\omega}_y$. For certain values of $\Omega$
this effect can  destabilize  the system. Physically
one should impose the conditions $\tilde{\omega}^2_x > 0$, 
$\tilde{\omega}^2_y > 0$ to ensure the normalizability of the density.

The equation of continuity (\ref{Rcontinuity}), which at 
equilibrium takes the form 
$({\bf v} - {\boldsymbol\Omega}\times {\bf r})\cdot{\boldsymbol\nabla}
\rho({\bf r}) = 0$, yields the following expression for $\alpha$
\begin{equation}
\alpha = -\Omega\left({\tilde{\omega}^2_x -\tilde{\omega}^2_y \over 
\tilde{\omega}^2_x +\tilde{\omega}^2_y}\right)
\label{alphaeq}
\end{equation}
in terms of $\Omega$ and of the effective frequencies $\tilde{\omega}_x, 
\tilde{\omega}_y$. 
From  (\ref{alphaeq}) and (\ref{radiimu}) one finds that  the
expectation value  of the angular momentum 
\begin{equation}
\langle L_z\rangle =
M\!\int\!\!\big({\bf r}\times{\bf v}\big)_z\,n({\bf r})\,d{\bf r} \equiv 
\Omega \Theta 
\label{L}
\end{equation} 
is always fixed by the irrotational value 
$\Theta =NM (\langle x^2-y^2\rangle)^2/\langle x^2 + y^2\rangle$ of
the moment of inertia \cite{inerzia}. In terms of the effective frequencies 
$\tilde{\omega_x}$ and $\tilde{\omega_y}$ the ratio between $\Theta$
and the classical rigid value 
$\Theta_{\rm rig} = NM \langle x^2 + y^2\rangle $ takes the simple expression
\begin{equation}
\frac{\Theta}{\Theta_{\rm rig}} =  
\left(\frac{\tilde{\omega}^2_x-\tilde{\omega}^2_y}{
\tilde{\omega}^2_x+\tilde{\omega}^2_y}\right)^2\;.
\label{Lz}
\end{equation}
Notice that both  $\Theta$ and $\Theta_{\rm rig}$  depend on the value of 
$\Omega$ since the square radii $\langle x^2\rangle$ and $\langle y^2\rangle$ 
are modified by the rotation. 

Another useful quantity to calculate is the release energy 
$E_{\rm rel} = E_{\rm kin}+E_{\rm int}$ giving the energy of the
system after switching off the confining trap. 
This quantity can be extracted from time of flight measurements 
on the expanding cloud. In non rotating condensates it
coincides with the interaction energy if one works in the
Thomas-Fermi regime. In the presence of the rotation the kinetic energy 
$E_{\rm kin}=M\int\!d{\mathbf r}\,n({\mathbf r})\,v^2/2$ 
cannot be instead neglected. By using the virial identity \cite{RMP}
 $2E_{\rm kin} - 2E_{\rm ho} + 3E_{\rm int} =0$, 
where $E_{\rm ho}$ is the expectation value of the oscillator potential, 
and noting that $E_{\rm int}=(2/7)N\tilde{\mu}$, one finds the result
\begin{equation}
\frac{E_{\rm rel}}{N}=\frac{\tilde{\mu}}{7}
\left(\frac{\omega_x^2}{\tilde{\omega}^2_x}+
\frac{\omega_y^2}{\tilde{\omega}^2_y}\right)\;.
\label{release}
\end{equation}

Let us now discuss the explicit behaviour of the stationary solutions of the
hydrodynamic equations (\ref{Rcontinuity})-(\ref{Rveq}).
By inserting expressions (\ref{tildex}) and (\ref{tildey}) into 
Eq.(\ref{alphaeq}) one finds the following third order equation for
$\alpha$ \cite{Recati}:
\begin{equation}
2\alpha^3+\alpha({\omega}^2_x + \omega_y^2 -4 {\Omega}^2)+ 
\Omega({\omega}^2_x-\omega_y^2) = 0\;.
\label{alpha3}
\end{equation}
Depending on the value of $\Omega$ and of the deformation 
\begin{equation}
\epsilon = \frac{\omega^2_x-\omega^2_y}{\omega^2_x+\omega^2_y}
\label{epsilon}
\end{equation}
of the trap, one can find either 1 or 3 solutions, derivable in
analytic form. As already anticipated, the physical
solutions should satisfy the additional requirements 
$\tilde{\omega}^2_x > 0$ and $\tilde{\omega}^2_y > 0$, which ensure
the normalizability of the density and rule out some of the
solutions of (\ref{alpha3}). The resulting phase diagram
is reported in Fig. 1 where, in the plane $\Omega-\epsilon$,
we show explicitly the regions characterized by 0,1,2 and 3 
solutions (we assume hereafter $\epsilon >0$). The solid curve, given by 
\begin{equation}
\frac{\epsilon^2\Omega^2}{\omega_x^2+\omega_y^2}+\frac{2}{27}
\left(1-4\frac{\Omega^2}{\omega_x^2+\omega_y^2}\right)^3=0\;,
\label{separation}
\end{equation}
divides the plane in two parts: on the left hand side
Eq. (\ref{alpha3}) admits only one solution, on the right hand side it
has three solutions. The dotted lines are the curves 
$\Omega = \omega_y=\omega_x\sqrt{(1-\epsilon)/(1+\epsilon)}$
and $\Omega=\omega_x$. If $\epsilon < 0.2$ 
one can find (see Fig. 1) 3 stationary solutions, by properly
choosing the value of $\Omega$ \cite{N1}. 
It is worth noticing that the phase diagram of Fig. 1 
differs from the one derivable in the non-interacting Bose gas
confined by the same harmonic trap. 
In this case the density profile in the rotating frame has the
Gaussian shape $\rho({\mathbf r}) =N\left(M\tilde\omega_{\rm ho}/ 
\pi\hslash\right)^{3/2}
\exp{\left[-M(\tilde{\omega}_xx^2+\tilde{\omega}_yy^2+{\omega}_zz^2)/\hslash
\right]}$.
The renormalized frequencies still obey the Equations (\ref{tildex})
and (\ref{tildey}), but the relationship for
$\alpha$ takes the different form $\alpha= -\Omega
(\tilde{\omega}_x-\tilde{\omega}_y)/(\tilde{\omega}_x+\tilde{\omega}_y)$.
In the non interacting case one finds that only one stationary 
solution is available  for $\Omega<\omega_y$ and $\Omega>\omega_x$, 
while no solution exists in the interval $\omega_y < \Omega
<\omega_x$.

In Figs. 2 and 3 we show the stationary
solutions of Eq. (\ref{alpha3}) for $\alpha$ in two interesting cases:
$\epsilon = 0.1$ where we predict the occurrence of a window 
with 3 stationary solutions, and $\epsilon =0.5$ where one has a
maximum of 2 stationary solutions satisfying the normalizability conditions
$\tilde{\omega}^2_x > 0$ and $\tilde{\omega}^2_y > 0$. 
The dashed-dotted line corresponds to the stationary solution of the non
interacting gas. In both Figs. 2 and 3 one
identifies two branches, hereafter called normal and overcritical
branches.

\begin{paragraph}{Normal branch:} 
this branch starts at $\Omega =0$. The linear dependence at small
$\Omega$ is given by $\alpha =-\Omega\epsilon$. 
By increasing $\Omega$ the square radius $\langle y^2\rangle$
increases and eventually diverges at $\Omega = \omega_y$ where 
$\tilde{\omega}_y \to 0$ and the branch has its end \cite{N2}. 
Also the angular momentum and the release energy diverge at
$\Omega=\omega_y$. Notice that when $\tilde{\omega}_y \to 0$
the moment of inertia takes the rigid value since $\langle
x^2\rangle\ll\langle y^2\rangle$ (see Eq.(\ref{Lz})). 
\end{paragraph}

\begin{paragraph}{Overcritical branch:}
this branch starts at $\Omega = +\infty$ where $\alpha$
behaves like $\alpha = (\omega_x^2-\omega_y^2)/4\Omega$.
It is worth noticing that in this limit both $\tilde{\omega}_x^2$ and 
$\tilde{\omega}_y^2$ approach the value $(\omega_x^2+\omega_y^2)/2$ and
therefore the shape of the density profile becomes symmetric despite
the asymmetry of the confining trap. In the
same limit the angular momentum tends to zero while the release energy
approaches the finite value $E_{\rm rel} =(2/7)\tilde{\mu}_\infty$, where 
$\tilde{\mu}_\infty$ is the chemical potential (\ref{mu}) with
$\tilde{\omega}_x^2=\tilde{\omega}_y^2=(\omega_x^2+\omega_y^2)/2$.
In the overcritical branch the deformation of the cloud 
takes a sign opposite to the one of the trap. 
This branch exhibits a back-bending at a value of $\Omega$ which
is smaller than $\omega_x$, but can be higher or smaller than
$\omega_y$, depending on whether the value
of $\epsilon$ is larger or smaller than $0.2$ 
(see Figs. 2 and 3). 
In both cases this branch ends, after the back-bending,
at the value $\Omega=\omega_x$, where $\tilde{\omega}_x \to 0$ and  
$\langle x^2\rangle$, $\langle L_z\rangle$ and $E_{\rm rel}$ diverge.
\end{paragraph}

It is also useful to discuss the instructive case $\epsilon\to 0$
corresponding to symmetric trapping in the $x$-$y$ plane
($\omega_x=\omega_y$).  
In this case one finds a solution  with $\alpha=0$ for
$\Omega <\omega_x/\sqrt{2}$ \cite{nota2}. 
For higher frequencies three solutions appear: the first one
still corresponds to a non-rotating configuration ($\alpha =0$), while two
solutions, given respectively by $\alpha =
\pm\sqrt{2\Omega^2-\omega_x^2}$, correspond to rotating deformed 
configurations.
The existence of these solutions, which break the original symmetry of
the Hamiltonian, is the analog of the
bifurcation from the axisymmetric Maclaurin to the triaxial Jacobi
ellipsoids for rotating classical fluids \cite{Chandrasekhar}. 
It is worth pointing out that the existence of the bifurcation is the
consequence of two-body interactions and is absent in the
non-interacting Bose gas.
When $\epsilon$ is slightly different from zero (and positive) the two
solutions are no longer degenerate, the one with $\alpha <0$ having
the lowest energy.

The existence of stationary solutions in the rotating frame raises the
important question of  stability. Actually, one should distinguish
between thermodynamic and dynamical instability.
The former corresponds to the absence of thermodynamic
equilibrium and  its signature, at zero temperature, 
 is given by the existence
of excitations with negative energy \cite{nota2}. The latter is instead
associated with the decay of the initial configuration due to
interaction effects and is in general revealed by the
appearance of excitations with complex energy.
Configurations characterized by thermodynamic instability 
are destabilized only in the presence of dissipative processes.

Let us first discuss the stability with respect to the center of mass motion. 
In the presence of harmonic trapping the corresponding equations of
motion, in the rotating frame, take the classical form (rotating
Blackburn's pendulum, \cite{Lamb}) and are not affected by interatomic forces. 
Their solutions obey the dispersion law
\begin{equation}
\omega^2=\frac{1}{2}\left[\omega_x^2+\omega_y^2+2\Omega^2\pm
\sqrt{(\omega_x^2-\omega_y^2)^2+8\Omega^2(\omega_x^2+\omega_y^2)}\right]\;,
\end{equation}
and are dynamically stable ($\omega^2 >0$) for $\Omega <\omega_y$ and
$\Omega >\omega_x$. So the requirement that the dipole oscillation be
dynamically stable excludes the region between the dotted
lines in Figs. 1, 2 and 3.
Notice that this is the same region where the Schroedinger equation
for the non-interacting Bose gas  has no stationary solutions in the
rotating frame.

We have further explored the conditions of stability by
studying the quadrupole oscillations of the condensate around the
equilibrium configuration in the rotating frame. The calculation is
derivable by linearizing the equations of motion (\ref{Rcontinuity})
and (\ref{Rveq}), with the choice
\begin{align}
\delta\rho ({\mathbf r})&=a_0+a_xx^2+a_yy^2+a_zz^2+a_{xy}xy
\label{deltarho}\\
\delta{\mathbf v}({\mathbf r})&={\boldsymbol{\nabla}}
(\alpha_xx^2+\alpha_yy^2+\alpha_zz^2+\alpha_{xy}xy)\;,
\label{deltav}
\end{align} 
for the fluctuations of the density and of the velocity field where the 
coefficients $a_i$ and $\alpha_i$ depend on time.
The results of the analysis show that the window of dynamical
instability for the quadrupole oscillations is different from the 
one of the dipole. 
In particular the normal branch is always dynamically stable, while the
overcritical branch is stable only in its lower part since, after the
back-bending, where d$\alpha/$d$\Omega >0$, one of the quadrupole
frequencies becomes purely imaginary
($\omega^2 <0$). We have investigated the stability of the
quadrupole excitations also in the limiting case $\epsilon =0$. 
In this case the upper and lower branches
$\alpha=\pm\sqrt{2\Omega^2-\omega_x^2}$ give rise to a vanishing value
for one of the quadrupole frequencies, the others being always real. 
This vanishing solution corresponds to the rotation of the system in the $x-y$
plane and
reflects the rotational symmetry of the Hamiltonian. 

We stress again that 
the solutions discussed in this paper correspond, in general, to
metastable configurations. For example it is well known
that vortical configurations can become energetically favorable for
relatively small values of $\Omega$.
Also excitations with higher multipolarities can become
energetically favorable for some values of the angular velocity
\cite{Muntsa}. In this work we assume that at very low temperature,
where collisions are rare, the solutions (\ref{virr})-(\ref{Rdensity})
can survive, at least for useful time intervals,  also in conditions
of thermodynamic instability.  

Let us finally briefly discuss the experimental possibility
of realizing the  rotations described above. The normal branch, 
characterized by the  huge increase of the size in the $y$ direction
and of the release energy when $\Omega\to\omega_y$, could
be in principle generated by an adiabatic increase of the
angular velocity, starting from a cold condensate \cite{Recati}.  
The transition from the normal to the overcritical branch cannot be instead
realized in a continuous way and the  only way to realize the
exotic configurations of this branch is to engineer the proper conditions 
of dynamical equilibrium, by building up both the proper phase of the
order parameter and the shape of the density profile.

We are indebted to S. Vitale for introducing us to the problem
of the overcritical rotations. Fruitful discussions with E.~Cornell,
J.~Dalibard, L.~Pitaevskii and G.~Shlyapnikov are also acknowledged.  
This work has been supported by the Ministero della Ricerca Scientifica e
Tecnologica (MURST).





\newpage
\begin{figure}
\begin{center}
\input{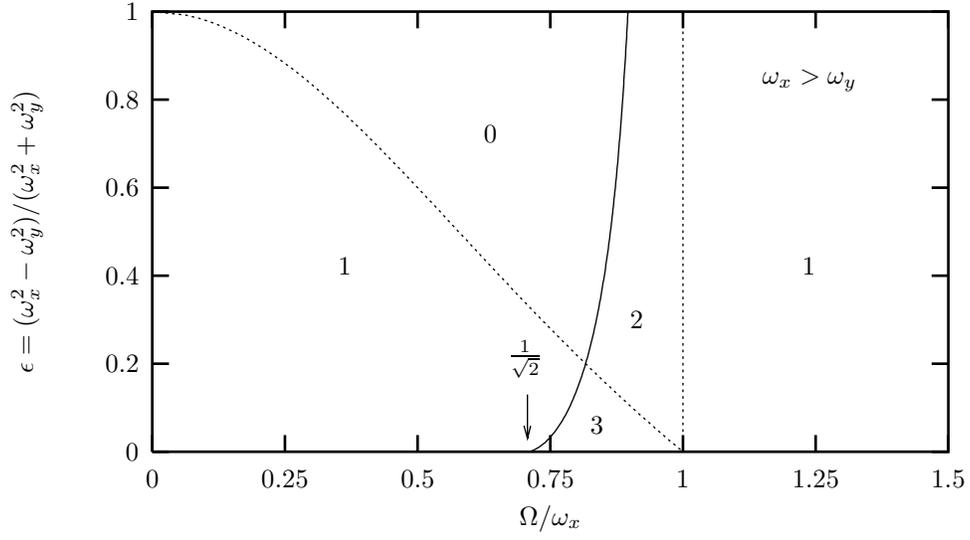}
\caption{Phase diagram representing the stationary
solutions of Eq. (\ref{alpha3}) (see text). 
The dotted lines are the curves $\Omega=\omega_y$ and 
$\Omega=\omega_x$ respectively. 
The full line is given by Eq. (\ref{separation}).}
\label{fig_diagram}
\end{center}
\end{figure}

\begin{figure}
\begin{center}
\input{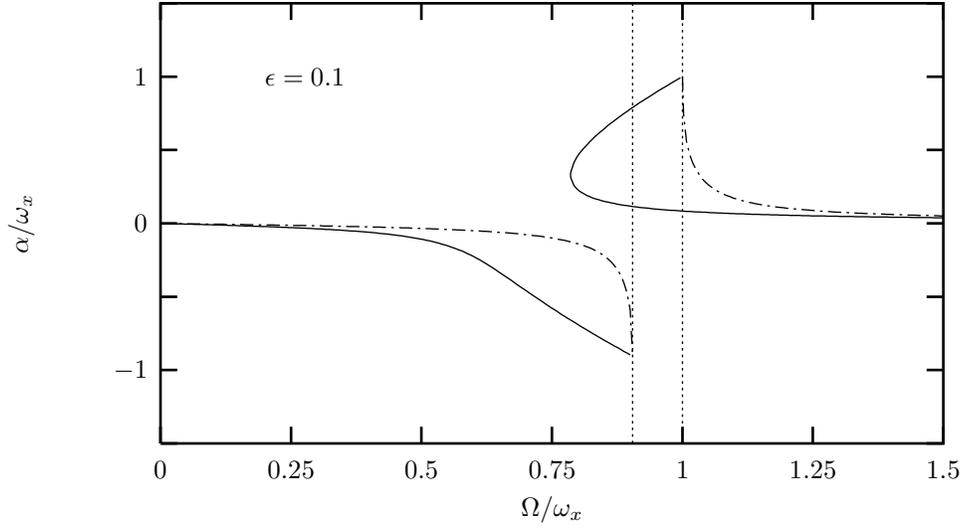}
\caption{Stationary solutions of Eq. (\ref{alpha3}) (solid lines) as a
function of $\Omega$, for $\epsilon =0.1$. 
The dashed-dotted curves are the stationary solutions of
the non-interacting Bose gas. The dotted straight lines correspond to 
$\Omega=\omega_y$ and $\Omega=\omega_x$ respectively.} 
\label{fig_e01}
\end{center}
\end{figure}

\begin{figure}
\begin{center}
\input{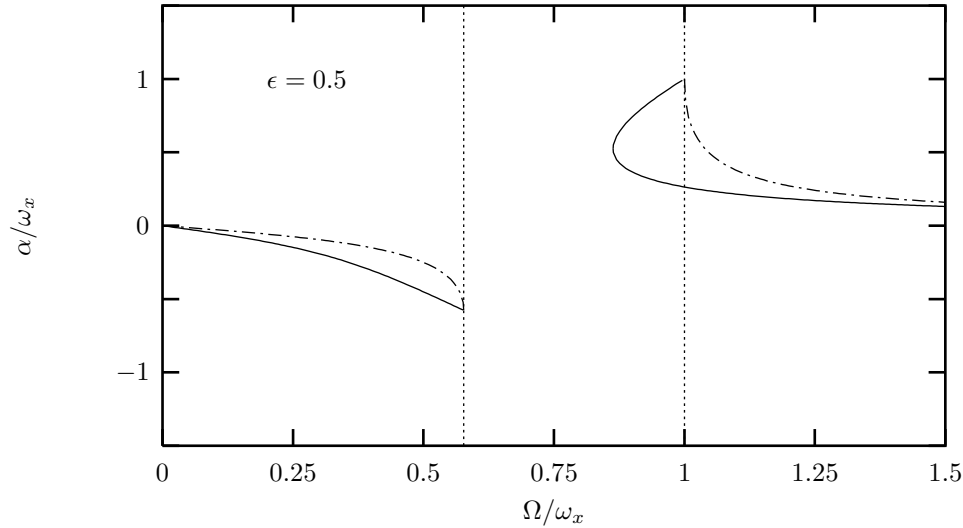}
\caption{As in Fig. 2, for $\epsilon =0.5$.}
\label{fig_e05}
\end{center}
\end{figure}


\begin{thebibliography}{99}
\bibitem{nota}
This choice, which is associated with a quadrupolar shape of the
equilibrium density (see Eq. (\ref{Rdensity})), corresponds to one of
the possible stationary solutions of the equations of motion.
\bibitem{N2} 
Near the end point the conditions of applicability of the Thomas-Fermi 
approximation are no longer valid and the solution of the 
problem should be obtained starting from the Gross-Pitaevskii equation.
\bibitem{nota2}
The value $\Omega =\omega_x/\sqrt{2}$ corresponds to the value of the
angular velocity for which the energy $\hbar\omega
=\hbar(\sqrt{2}\omega_x-2\Omega)$ of the $L_z=+2$ quadrupole oscillation,
relative to an axisymmetric configuration, vanishes in the rotating
frame \cite{SS}. For larger $\Omega$ the excitation energy becomes negative.
\end{thebibliography}
\end{document}